\documentstyle[12pt]{article}
\renewcommand{\baselinestretch}{1.2}
\def\head{\vspace{-7cm}\begin{flushright} DESY 93-192\\[-2mm]
                                          December 1993\\[1cm]
                        \end{flushright}}
\newcommand{\ba}{\begin{array}}
\newcommand{\ea}{\end{array}}
\newcommand{\bd}{\begin{displaymath}}
\newcommand{\ed}{\end{displaymath}}
\newcommand{\be}{\begin{equation}}
\newcommand{\ee}{\end{equation}}
\newcommand{\bea}{\begin{eqnarray}}
\newcommand{\eea}{\end{eqnarray}}

\def\qb{\bar{q}}
\def\ub{\bar{u}}
\def\db{\bar{d}}
\def\cb{\bar{c}}
\def\sb{\bar{s}}

\def\i{i}                  
\def\to{\rightarrow}       
\def\Im{Im}
\def\Re{Re}
\def\ln{\mbox{$\ell n$}}   
\def\sin{\mbox{ sin}}
\def\cos{\mbox{ cos}}
\def\Sum{\displaystyle\sum}
\def\eff{{\rm eff}}
\def\Heff{{\cal H}_{\rm eff}}
\def\ce{c^{\eff}}
\def\me{{\bf m}}
\def\r{{\bf r}}

\def\JPsi{J/\Psi}

\def\eq{eq.~}

\def\bra{\langle}
\def\ket{\rangle}

\def\a{\alpha}
\def\b{\beta}
\def\g{\gamma}
\def\d{\delta}

\def\l{\lambda}
\def\m{\mu}

\def\G{\Gamma}
\def\D{\Delta}

\begin{document}
\head
\begin{center} \begin{Large} \begin{bf}
CP Violation and Strong Phases from Penguins in $\bf B^{\pm}\rightarrow VV$
Decays
\end{bf} \end{Large} \end{center}
\vspace{1cm}
\begin{center}
    G.\ Kramer$^a$,
    W.\ F.\ Palmer$^b$\footnote{Supported in part by the US
            Department of Energy under contract DOE/ER/01545-605.},
    H.\ Simma$^c$\\
      \vspace{0.3cm}
        $^a$II. Institut f\"ur Theoretische Physik\footnote{Supported by
             Bundesministerium f\"ur Forschung und Technologie,
             05\,6\,HH\,93P(5), Bonn, FRG.}\\ der Universit\"at Hamburg,\\
        D--22761 Hamburg, Germany\\
  \vspace{0.3cm}
        $^b$Department of Physics, The Ohio State University, \\
        Columbus, Ohio 43210, USA\\
  \vspace{0.3cm}
        $^c$Deutsches Elektronen Synchrotron DESY\\
        D--22603 Hamburg, Germany\\
        \end{center}
  \vspace{1cm}

\noindent {\bf Abstract}\\
\parbox[t]{\textwidth}{
We calculate direct CP-violating observables in charged
$B\to VV$ decays arising from the interference of amplitudes with different
strong and CKM phases. The perturbative strong phases develop at order
$\alpha_s$ from absorptive parts of one-loop matrix elements of the
next-to-leading logarithm corrected effective Hamiltonian. CPT
constraints are maintained.  Based on this model, we find that
partial rate asymmetries between charge conjugate $B^{\pm}$ decays
can be as high as 15-30\% for certain channels with branching
ratios in the $10^{-6}$ range.  The small values of the coefficients
of angular correlations, which we calculated previously to be of
order $10^{-2}$, are not significantly degraded by the strong
phases.  The charge asymmetries of rates and angular distributions
would provide unambiguous evidence for direct CP violation.
}
\newpage

\section{Introduction}

So far CP violation \cite{Jarl} has been detected only in processes
related to $K^0 - \bar K^0$ mixing \cite{CCFT} but considerable
efforts are being made to investigate it in $B$ decays. While the most
promising proposal for observing CP violation in the $B$ system involves the
mixing between neutral $B$ mesons \cite{bigi}, the particular interest
in decays of charged $B$ mesons lies in their possibilities
for establishing the detailed nature of CP violation.
Since charged $B$ mesons can not mix, a measurement of a CP violating
observable in these decays would be a clear sign for {\it direct} CP
violation, which has been searched for in $K$ decays with indefinite
success as the measurements of $\epsilon'/\epsilon$ do not yet
exclude a zero value \cite{BP}.

In non-leptonic charged $B$ decays two main categories of CP-violating
observables can be investigated: First, rate asymmetries \cite{BSS,GH}, \be
a_{CP} = \frac{\Gamma - \bar \Gamma}{\Gamma + \bar \Gamma} \ ,\ee
where $\Gamma$ and $\bar{\Gamma}$ are the (partial) rates of the decay
and its charge (C) conjugate, and second, azimuthal angular correlations
\cite{Val,KP}.

The rate asymmetries occur even for spinless final
states and require both weak {\it and} strong phase differences in
interfering amplitudes.  The weak phase differences arise from the
superposition of amplitudes from various penguin diagrams and the
usual W-exchange (if contributing).
The strong phase is generated by final state interactions. At the quark
level these strong interaction effects
can be modeled by absorptive parts of perturbative penguin
diagrams (hard final state interactions) \cite{BSS} while predictions at
the non-perturbative hadronic level are of course extremely difficult
(soft final state interactions).
Clearly we can not exclude that the weak transition matrix elements
receive phases originating from soft final state interactions (resonances)
between the produced vector particles. However, since the mass
of the $B$ is far above the usual resonance region, we expect these
phase shifts to be small.

It is clear that a significant
contribution of penguin diagrams, and hence of the CKM \cite{CKM} phase
differences, is an exceptional case and requires either the absence or
a strong CKM suppression of the tree contributions (as e.g. in charmless
$b \rightarrow s$ transitions).

The rate asymmetries for exclusive two-body decays into pseudoscalars
have been estimated by several authors using either the model of Bauer,
Stech and Wirbel \cite{BSW} (BSW) based on wave functions in the infinite
momentum frame, or the perturbative methods developed by Brodsky et. al.
\cite{BLS}. The rate asymmetries $a_{CP}$ can be quite large
 (of the order $a_{CP} \sim 0.1)$
for some of the final states.  However, the corresponding branching
fractions of these decays are quite small, ranging from
$10^{-6}$ (estimates with the BSW model \cite{BSW}) to $10^{-7}$
(estimates with the Brodsky-Lepage model \cite{BLS,SW,Flei}).
The magnitude of $a_{CP}^2 \times BR$ is therefore of the order of
$10^{-9} - 10^{-7}$.

The second category of CP-violating observables
involves the decay of the $B$ meson into two vector particles
$B \rightarrow V_1 V_2$ with subsequent decays of $V_1$ and $V_2$
\cite{Val,KP}. (In the following $B$ will always denote the $B^-$ meson
and $\bar B$ its antiparticle.)
By analyzing the azimuthal dependence of the vector meson decay products
one can then isolate CP odd quantities.  The advantage of this method
is that the CP violating terms occur even when there are no strong phase
differences between the interfering weak amplitudes. On the other
hand these coefficients in the azimuthal correlation are also present
when the CP-violating weak phase differences vanish. By measuring these
coefficients in charge conjugate $B^\pm$ decays one has the possibility
to disentangle the effects of strong and weak phases \cite{KP}.

So far the ``CP-odd'' azimuthal angular coefficients have been calculated
under the assumption that the strong phase differences vanish \cite{KP}.
These results give us an estimate of the effect we could expect from the
CP-violating phase. Since strong phase differences are present, at least
due to the hard final state interactions which lead to the rate asymmetries
through the Bander-Silverman-Soni mechanism \cite{BSS}, we should include
these strong phases also in the calculation of the ``CP-odd'' angular
coefficients.

The two categories for detecting direct CP violation are therefore
complementary. The rate asymmetries occur only when non-vanishing strong
phase differences are present but need no angular information.  The azimuthal
angular correlation terms need no strong phase differences but require joint
angular distribution measurements of the decay products of the $B$ mesons
\cite{Bia}.

Although the explicit form of these angular distributions depends on the spins
of the decay products of the decaying vector mesons $V_1$ and $V_2$, two
formulas are sufficient to describe a general B$\to$VV decay \cite{KP}.
For instance, the angular distribution for the cascade decay
$B^-\to K^{*-} \rho^0 \rightarrow \left(K\pi\right)\left(\pi^+\pi^-\right)$
has the following form:
\bea
\frac{d^3\Gamma}{dcos\theta_1dcos\theta_2d\phi}
& \sim &
\frac{1}{4}
\frac{\G_T}{\G} \cdot
sin^2\theta_1\ sin^2\theta_2
+\frac{\G_L}{\G} \cdot
cos^2\theta_1\ cos^2\theta_2
\nonumber  \\[3mm]
&&+\frac{1}{4}\sin2\theta_1 \ sin2\theta_2 \lbrack
\a_1 \cdot
cos\phi
-
\b_1 \cdot
sin\phi
\rbrack
\label{distribution} \\[3mm]
&&+\frac{1}{2}sin^2\theta_1sin^2\theta_2\ \lbrack
\a_2 \cdot
cos2\phi\
-
\b_2 \cdot
sin2\phi\
\rbrack
\ . \nonumber \eea

In \eq(\ref{distribution}) $\theta_1$ is the polar angle of the $K$ in the
rest system of the $K^*$ with respect to the helicity axis. Similarly
$\theta_2$ and $\phi$ are the polar and azimuthal angle of the $\pi^+$ in
the $\rho^0$ rest system with respect to the helicity axis of the $\rho^0$;
i.e. $\phi$ is the angle between the planes of the two decays
$K^{*-}\to K\pi$ and $\rho^0 \to \pi^+\pi^-$.

The decay distribution is parameterized by the coefficients:
\be\ba{llllll}
\frac{\Gamma_T}{\Gamma} & = &
\frac{\vert H_{+1}\vert ^2 + \vert H_{-1}\vert ^2}
{\vert H_0\vert ^2 + \vert H_{+1}\vert ^2 + \vert H_{-1}\vert ^2}
\ \ \ \ \ &
\frac{\Gamma_L}{ \Gamma} & = &
\frac{\vert H_0\vert ^2}{\vert H_0\vert ^2 + \vert H_{+1}\vert ^2 +\vert H_{-1}
\vert ^2} \\[5mm]
\alpha_1 & = &
\frac{Re\left (H_{+1}H_0^\ast + H_{-1}H_0^\ast\right )}
{\vert H_0\vert ^2 + \vert H_{+1}\vert ^2 + \vert H_{-1}\vert ^2} &
\beta_1 & = &\frac{Im\left (H_{+1} H_0^\ast- H_{-1}H_0^\ast\right)}
{\vert H_0 \vert ^2+ \vert H_{+1} \vert ^2 + \vert H_{-1} \vert ^2 }\\[5mm]
\alpha_2 & = & \frac{Re\left(H_{+1}H_{-1}^\ast\right)}
{\vert H_0\vert ^2 + \vert   H_{+1}\vert ^2 + \vert H_{-1}\vert ^2} &
\beta_2 & = & \frac{Im\left (H_{+1} H_{-1}^\ast\right)}
{\vert H_0\vert ^2 +\vert H_{+1}\vert ^2 + \vert H_{-1}\vert ^2} \\
\ea \label{parameters}\ee
where
$H_\lambda = \bra V_1(\lambda)V_2(\lambda)\vert \Heff \vert  \bar{B}\ket $
are the helicity amplitudes ($\lambda = 0, \pm 1$).
Clearly \eq(\ref{distribution}) can be used also for all other decays
where $V_1$ and $V_2$ decay into two pseudo scalar mesons. The other decay
distribution, e.g. for  $B\rightarrow D_s^\ast D^\ast$ with
$D_s^\ast\rightarrow D_s\gamma$ and $D^\ast\rightarrow D\pi$
can be found in ref.~\cite{KP}, where further examples are discussed.

In our previous work, we have calculated the six angular coefficients
and the branching ratios for 36 decays $B\to V_1 V_2$ with neutral and
charged $B$ mesons.  Penguin contributions were taken into account only
through leading-log short distance QCD effects and all strong phases were
neglected.
Non-vanishing ``CP-odd'' azimuthal angular correlations, i.e. coefficients
$\b_1$ and $\b_2$ occurred only in the decays $B^- \to K^{*-}\omega$,
$K^{*-}\rho^0$, $\omega\rho^-$ (and the corresponding decays of $\bar{B^0}$).
In this work, we shall present results including strong phases from
penguin diagram contributions to the matrix elements. We consider
the rate asymmetries $a_{\rm CP}$ and the decay parameters from
\eq(\ref{parameters}) for $B^-$ and $B^+$ decays, and investigate
the influence of the strong phases on $\b_{1,2}$. In addition, we
base our treatment on the next-to-leading logarithmic short distance
corrections evaluated by Buras et al. \cite{Buras1}, which is mandatory if
one wants to systematically take into account the complete
$O(\a_s)$ penguin matrix
elements. We include also some pure penguin modes which are of interest
for the detection of CP effects via the rate asymmetry $a_{\rm CP}$, and
we give estimates of their branching ratios.

The remainder of this paper is organized as follows.
In section~2 we describe the effective weak Hamiltonian and the evaluation
of the hadronic matrix elements. The CP-violating observables are discussed
in section~3. The final results for the angular correlations and rate
differences are discussed in section~4. Formulae for the matrix elements and
some technical details about CPT cancellations can be found in the appendices.

\section{The effective Hamiltonian}
\subsection{Short distance QCD corrections}
For calculations of CP-violating observables it is most convenient to
split the effective weak Hamiltonian into two pieces, one proportional to
$v_u\equiv V_{ub}V_{us}^\ast$ (or $V_{ub}V_{ud}^\ast$ in the case of
$b\to d$ transitions) and the other one proportional to
$v_c\equiv V_{cb}V_{cs}^\ast$ (or $V_{cb}V_{cd}^\ast$ correspondingly),
\be \Heff = 4 \frac{G_F}{\sqrt{2}} \left( v_u \Heff^{(u)} + v_c \Heff^{(c)}
\right) \ . \ee
The two terms ($q = u,c$)
\bd \Heff^{(q)} = \sum_i c_i(\mu) \cdot O_i^{(q)}\ , \ed
differ only by the quark content of the local operators,
and for our purposes it is sufficient to consider only
the following four-quark operators \cite{GW}:
\be\ba{llllll}
O_1^{(q)} & = & \sb_\a \g^\mu L q_\b \cdot \qb_\b \g_\mu L b_\a \ , \ \ &
O_2^{(q)} & = & \sb_\a \g^\mu L q_\a \cdot \qb_\b \g_\mu L b_\b \ , \\
O_3 & = & \sb_\a \g^\mu L b_\a \cdot \Sum_{q'}\qb_\b'\g_\mu L q_\b'\ , \ \  &
O_4 & = & \sb_\a \g^\mu L b_\b \cdot \Sum_{q'}\qb_\b'\g_\mu L q_\a'\ , \\
O_5 & = & \sb_\a \g^\mu L b_\a \cdot \Sum_{q'}\qb_\b'\g_\mu R q_\b'\ , \ \  &
O_6 & = & \sb_\a \g^\mu L b_\b \cdot \Sum_{q'}\qb_\b'\g_\mu R q_\a'\ .
\label{operators}
\ea \ee
where $L$ and $R$ are the left- and right-handed projection operators.
The operators $O_3,\ldots,O_6$ arise from (QCD) penguin diagrams which
enter at order $\a_s$ in the initial values of the coefficients,
\bd c_i(M_W) = \left\{
\ba{ll} 1+O(\a_s) & (i=2)\\ O(\a_s) & ({\rm otherwise}) \ea \right. \ , \ed
or through operator mixing during the renormalization group
summation of short distance QCD corrections. The renormalization group
evolution from $\mu \approx M_W$ to $\mu \approx m_b$ has been evaluated
in next-to-leading logarithmic (NLL) precision by Buras et al. \cite{Buras1}.
These authors also demonstrated how the $O(\a_s)$ renormalization scheme
dependence can be isolated in terms of a matrix $\r_{ji}$ by writing
\be c_j(\mu) = \sum_i \bar{c}_i(\mu)
 \left[ \d_{ij} - \frac{\a_s(\mu)}{4\pi} \r_{ij}\right] \ , \ee
where the coefficients $\cb_j$ are scheme independent at this order.
The numerical values for $\Lambda^{(4)}_{\overline{MS}} = 350$ MeV
\footnote{
   This value of $\Lambda^{(4)}_{\overline{MS}}$ translates to
   $\Lambda^{(5)}_{\overline{MS}} \approx 250 MeV$, which is about
   the value from a recent compilation of G. Altarelli
   ($\Lambda_{\overline{MS}}^{(5)} = 240\pm90 MeV$ \cite{Alt})
}, $m_t = 150$ GeV and $\mu = m_b = 4.8$GeV are \cite{Buras1}
\be\ba{llllll}
\cb_1 & = & - 0.324\ , \ \ & \cb_2 & = & 1.151\ , \\
\cb_3 & = & 0.017\ , \ \ & \cb_4 & = & -0.038\ , \\
\cb_5 & = & 0.011\ , \ \ & \cb_6 & = & -0.047\ .
\ea\ee

Contributions from the color magnetic moment operator
\bd O_g = \frac{g_s}{16 \pi^2}
\cdot \Psi \sigma _{\mu \nu}\left( m_{b}R+m_{s}L\right)
\frac{\lambda ^{a}}{2}\Psi
\; G^{\mu \nu}_{a} \ , \ed
with a coefficient of the order $-0.15$,
will allways be neglected in the following, because already its tree level
matrix elements are suppressed by a factor $\alpha_s/4\pi$ and it cannot
provide interesting absorptive parts in the decays considered here.

\subsection{Quark-level matrix elements}
Working at NLL precision, it is consistent -- and necessary in order to
cancel the scheme dependence from the renormalization group evolution
 -- to treat the matrix
elements of $\Heff$ at the one-loop level. These one-loop matrix elements
can be rewritten in terms of the tree-level matrix elements of the
effective operators, and one obtains:
\be \bra sq'\qb'\vert \Heff^{(q)}\vert b\ket =
\sum_{i,j} c_i(\mu)
\left[ \d_{ij} + \frac{\a_s(\mu)}{4\pi} \me_{ij}(\mu,\ldots)\right]
\bra sq'\qb'\vert O_j^{(q)}\vert b\ket^{\rm tree} \ . \label{me1}\ee
The functions $\me_{ij}$ are determined by the corresponding renormalized
one-loop diagrams and depend in general on the scale $\mu$, on the quark
masses and momenta, {\it and} on the renormalization scheme. The various
one-loop diagrams can be grouped into two classes: {\it vertex-corrections},
where a gluon connects two of the outgoing quark lines (fig.~1a),
and {\it penguin} diagrams, where a quark-antiquark line closes a loop
and emits a gluon, which itself decays finally into a quark-antiquark pair
(fig.~1b).

When expressing the rhs of \eq(\ref{me1}) in terms of the renormalization
scheme independent coefficients $\cb_i$, the effective coefficients multiplying
the matrix elements $\bra sq'\qb'\vert O_j^{(q)}\vert b\ket^{\rm tree}$
become
\be c_j^\eff \equiv \cb_j + \frac{\a_s}{4\pi} \sum_i \cb_i \cdot
\left( \me_{ij}-\r_{ij} \right) \ . \ee
The renormalization scheme dependence, which is present in $\me_{ij}$
and $\r_{ij}$, explicitly cancels in the combination $\me_{ij}-\r_{ij}$.
This reflects the familiar fact that one-loop matrix elements have to be
included in order to compensate for the order $\a_s$ renormalization
scheme dependence which enters through the coefficients $c_i$ in $\Heff$
(generated by the NLL renormalization group evolution).
For instance, the scheme dependence in the coefficients of the penguin
operators, which enter via $c_j \bra sq'\qb'\vert O_j^{(q)}\vert
b\ket^{\rm tree}$ for $j=3,4,5,6$ in \eq(\ref{me1}), is cancelled
by penguin-like matrix elements (fig.~1b) of the operators.

The effective coefficients $c_{1,2}^\eff$ receive contributions only from
{\it vertex-correction} diagrams, which will not be included in the following
(see the discussion in Section~3).
For a general $SU(N)$ color group the remaining effective coefficients can
be brought into the following form
\bea
c_3^\eff & = & \cb_3 - \frac{1}{2N} \frac{\a_s}{4\pi} (c_t + c_p)
     + \cdots \nonumber\\
c_4^\eff & = & \cb_4 + \frac{1}{2} \frac{\a_s}{4\pi} (c_t + c_p)
     + \cdots \nonumber\\
c_5^\eff & = & \cb_5 - \frac{1}{2N} \frac{\a_s}{4\pi} (c_t + c_p)
     + \cdots \nonumber\\
c_6^\eff & = & \cb_6 + \frac{1}{2} \frac{\a_s}{4\pi} (c_t + c_p)
    + \cdots \ , \label{c3456} \eea
where we have separated the contributions $c_t$ and $c_p$ from the ``tree''
operators $O_{1,2}$ and from the penguin operators $O_{3\cdots6}$,
respectively. The ellipses denote further contributions from
{\it vertex-correction} diagrams.

In addition to the contributions from penguin diagrams with insertions of
the tree operators $O^{(q)}_{1,2}$
\be
c_t = \cb_2 \cdot \left[\frac{10}{9}+\frac{2}{3} \ln \frac{m_q^2}{\mu^2}
- \Delta F_1\Bigl(\frac{k^2}{m_q^2}\Bigr) \right]\ , \label{ct} \ee
where $\Delta F_1$ is defined in appendix~A, we have evaluated
the penguin diagrams for the matrix elements of the penguin operators:
\bea
c_p & = & \cb_3 \cdot \left[
       \frac{280}{9}
       + \frac{2}{3} \ln \frac{m_s^2}{\mu^2}
       + \frac{2}{3} \ln \frac{m_b^2}{\mu^2}
       - \Delta F_1\Bigl(\frac{k^2}{m_s^2}\Bigr)
       - \Delta F_1\Bigl(\frac{k^2}{m_b^2}\Bigr) \right] \nonumber \\
& + & (\cb_4+\cb_6)\cdot \sum_{j=u,d,s,\ldots} \left[
       \frac{10}{9}+\frac{2}{3} \ln \frac{m_j^2}{\mu^2}
      - \Delta F_1\Bigl(\frac{k^2}{m_j^2}\Bigr) \right] \ , \label{cp} \eea

\subsection{Hadronic matrix elements in the BSW model}
In order to evaluate the hadronic matrix elements of $\Heff$ we represent
the helicity amplitudes in terms of three invariant amplitudes, $a , b$ and
$c$:
\be
H_\lambda \equiv \epsilon_{1\mu}(\lambda)^\ast \epsilon_{2\nu}(\lambda)^\ast
\left[ a g^{\mu\nu} + \frac{b}{m_1 m_2}~p_2^\mu p_1^\nu
    + \frac{ic}{m_1m_2} \epsilon ^{\mu\nu\alpha\beta} p_{1\alpha}  p_{2\beta}
\right] \ , \label{invamp}\ee
where $p_{1,2}$ and $m_{1,2}$ are the four-momenta and masses
of $V_{1,2}$, respectively.

The coefficients $a , b$ and $c$ have strong phases $\delta$ from
final state interactions (e.g. between the
two vector particles $V_1$ and $V_2$) and weak phases $\phi$
 originating from the CP violating phase in the  CKM  matrix.
In general, the invariant amplitudes are a sum of several interfering
amplitudes, $a_k$, $b_k$, and $c_k$, respectively, corresponding to,
for instance, various different isospin contributions.
Then the phase structure of $a,\ b$ and $c$ is:
\bea
a & = & \sum_{k} \vert a_k\vert e^{i\delta_{a_k} + i\phi_{a_k}} \nonumber \\
b & = & \sum_{k} \vert b_k\vert e^{i\delta_{b_k} + i\phi_{b_k}} \label{abc}\\
c & = & \sum_{k} \vert c_k\vert e^{i\delta_{c_k} + i\phi_{c_k}} \nonumber
\ . \eea

\def\bara{\tilde{a}}
\def\barb{\tilde{b}}
\def\barc{\tilde{c}}
The helicity amplitudes   $\bar H_\lambda$ for the decay of $\bar B \rightarrow
\bar V_1 \bar V_2$, where $\bar V_1$ and $\bar V_2$ are the antiparticles of
$V_1$ and $V_2$ respectively have the same decomposition as (12)
with  $a \rightarrow \bara, b \rightarrow \barb$ and $c \rightarrow -
\barc$. The amplitudes $\bara, \barb$
and $\barc$ have the analogous phase structure as above with the weak
phases changing sign $\phi_{a_k,b_k,c_k}\rightarrow - \phi_{a_k,b_k,c_k}$;
i.e. the $\bara_k $, $\barb_k $ and $\barc_k$ have the same strong
phase shifts and the opposite weak phase compared to $a_k$, $b_k$ and $c_k$.

{}From the decomposition \eq(\ref{invamp}) one finds the following relations
between the helicity amplitudes and the invariant amplitudes, $a, b, c$:
\be
H_{\pm 1} = a \pm c \sqrt{x^2-1} \ \ \ {\rm and} \ \ \
H_0 = -ax -b\left(x^2-1\right) \ , \ee
where
\bd x = \frac{p_1 p_2}{m_1 m_2} = \frac{m_B^2 - m_1^2 - m_2^2}{2 m_1 m_2}
\ . \ed
When there are no strong interaction phases, one has
$\bara = a^* ,\; \barb = b^*$ and $\barc = c^*$. Due to the
sign change in front of $\barc$ in $\bar H_\lambda$, we have in this case
\bd \bar H_{\pm1} = H{_{\mp1}^{\ast}}\ , \ \ \bar H_0 = H_{0}^\ast \ .\ed

To take into account long distance QCD effects which build up the
hadronic final states, we follow Bauer, Stech and Wirbel \cite{BSW}:
With the help of the factorization hypothesis \cite{Fak} the three-hadron
matrix elements are split into vacuum-meson and meson-meson matrix elements
of the quark currents entering in $O_1,\ldots,O_6$. In addition, OZI
suppressed form factors and annihilation terms are neglected.
In the BSW model, the meson-meson matrix elements of the currents are
evaluated by overlap integrals of the corresponding wave functions and
the dependence on the momentum transfer (which is equal to the mass of
the factorized meson) is modeled by a single-pole ansatz.
As a first approximation, this  calculational scheme
provides a reasonable method for estimating the relative size and
phase of the tree and penguin terms that give rise to the CP-violating
signals.

Concerning the QCD coefficients and how $1/N$ terms are treated, it is
well known \cite{Stone} that this model has problems accounting
for the decays with branching ratios which are proportional to $a_2^2$, where
\bd
a_2 = \bar{c}_1 + \frac{1}{N} \bar{c}_2 \ .\ed This is due to the fact
that $a_2$ has a rather small value $\vert a_2\vert \approx 0.06$ when
using the short-distance QCD corrected coefficients.  An analogous
effect is also known in nonleptonic D decays \cite{BSW}, and several
authors advocated a modified procedure to evaluate the factorized
amplitudes \cite{BSW,Buras2}: There, only terms which are dominant in
the $1/N$ expansion are taken into account.
Recently there has been much discussion in the literature concerning
these issues.  Some authors have argued that QCD sum rules validate
this procedure \cite{sum}.
As our model for evaluating the matrix elements of the weak Hamiltonian
we also choose this leading $1/N$ approximation and use the QCD corrected
coefficient functions $\cb_i$ given above.
We note that the terms proportional to $1/N$ in \eq(\ref{c3456}) must
then be dropped as well.

The strong phase shifts are generated in our model only by the
absorptive parts (hard final state interactions) of the quark-level
matrix elements of the effective Hamiltonian. Of course, when factorizing
the hadronic matrix elements, all information on the crucial value of the
momentum transfer $k^2$ of the gluon in the penguin diagram (fig.~1b) is
lost. While it has been attempted \cite{SW}
to model a more realistic momentum distribution by taking into account
the exchange of a hard gluon, we will use here for simplicity only a
fixed value of $k^2$. From simple two body kinematics \cite{Deshpande}
or from the investigations in ref.~\cite{SW} one expects $k^2$ to be
typically in the range
\be
\frac{m_b^2}{4}\stackrel{<}{\sim} k^2 \stackrel{<}{\sim}\frac{m_b^2}{2}\ .
\ee

\section{CP-violating observables}
In general, CP-violating observables require the
interference of two amplitudes with different weak phases, $\phi$, from
the CKM factors. To investigate the necessity of strong phases for the
observables in  $B\to VV$ decays (see section~1) and their interplay
with the weak phase factors, we decompose the amplitudes
into contributions proportional to $v_u$ and $v_c$
\be  A = v_u \cdot A^{(u)} + v_c \cdot A^{(c)} \ , \label{auc} \ee
where $A$ stands for a generic decay or helicity amplitude.
The amplitudes for the CP conjugate process, $A^{CP}$, are then
obtained from \eq(\ref{auc}) by replacing $v_q$ by $v_q^\ast$.

Observables which involve only the real parts of the interfering
amplitudes, like the decay rate or the parameters $\a_{1,2}$, can
signal CP violation only when one compares them with the corresponding
quantities of the charge conjugate decay channel, and when both,
non-vanishing weak phase differences {\it and} strong phase shifts,
$\d_u -\d_c$, are present. For instance, (defining
$v_q \equiv \vert v_q\vert e^{i\phi_q}$)
\be \G - \bar{\G} \sim \Im [v_u v_c^\ast] \cdot \Im [A_u A_c^\ast]
\sim \sin (\phi_u -\phi_c)\cdot \sin (\d_u -\d_c)\ . \label{delta} \ee

On the other hand, the decay parameters $\b_i$ ($i=1,2$) can have non-zero
values in the presence of either weak {\it or} strong phases alone.
Then, by comparison with the parameters $\bar{\b}_i$ ($\b_i^{CP}$) of
the C (CP) conjugate decay, one can, in principle, establish a weak
phase difference even for vanishing strong phases
\be
\b_i + \bar{\b}_i \ = \ \b_i - \b_i^{CP} \ \sim\
\Im [v_u v_c^\ast] \cdot \Re [A_u A_c^\ast]
\ \sim\ \sin (\phi_u -\phi_c)\cdot \cos (\d_u -\d_c) \ . \label{beta1}\ee
or measure the strong phase shifts even for negligible weak phases
\be
\b_i - \bar{\b}_i\ = \ \b_i + \b_i^{CP} \  \sim\
\Re [v_j v_k^\ast] \cdot \Im [A_j A_k^\ast]
\ \sim\ \cos(\phi_j -\phi_k) \cdot \sin (\d_j -\d_k) \ ,
\label{beta2} \ee
where $v_j$ and $v_k$ are not necessarily different.
In eqs.~(\ref{beta1}) and (\ref{beta2}) we have dropped
terms proportional to $\Im [v_u v_c^\ast] \cdot \Im [A_u A_c^\ast]$
which arise from the different denominators, $\G$ and $\bar\G$, in
$\b_i$ and $\bar\b_i$. Note also the relative sign between $\bar{\b}_i$
and $\b_i^{\rm CP}$ due to the parity reflection.

If there are no CP-violating weak phases then $\beta_i = -\bar \beta_i$,
$\alpha_i= \bar \alpha_i$, and $\Gamma = \bar{\Gamma}$,
while the absence of strong phases implies $\beta_i = \bar \beta_i$ (and,
of course, $\alpha_i = \bar \alpha_i$ and $\Gamma = \bar{\Gamma}$).
Interesting CP differences in the case $B\to VV$, which do not require
strong phases [see \eq(\ref{beta1})] and which are proportional to weak
phase differences, are [using the phase definitions of \eq(\ref{abc})]
\be Im(H_{+1}H^*_{-1} + \bar H_{+1} \bar H^*_{-1}) =
        - 4 \sqrt{x^2-1} \sum_{i,j} sin(\phi_{ai}-\phi_{cj}) \;
        cos(\delta_{ai}-\delta_{cj}) \vert a_i c_j\vert \ee
\noindent and
\bea
Im(H_{+1}H^*_{0} -H_{-1}H_0^*+ \bar H_{+1} \bar H_0^*
 - \bar H_{-1} \bar H_0^*) =\hspace{5cm} && \nonumber \\
-4(x^2-1)^{\frac{3}{2}} \sum_{i,j} sin(\phi_{ci}-\phi_{bj})\;
        cos(\delta_{ci}-\delta_{bj})\vert c_i b_j \vert &&\\
-4 x \sqrt{x^2-1}\sum_{i,j} sin(\phi_{ci}-\phi_{aj})\;
        cos(\delta_{ci}-\delta_{aj})\vert c_i a_j\vert &.&
\nonumber \eea
Here, $\bar H_\l$ are the amplitudes of the charge conjugated
process and the subscripts of the weak and strong phases refer
to different weak (or isospin, etc.) contributions.

The presence of strong phases is unambiguously demonstrated by a partial
rate asymmetry as well as angular correlations of the following kind:
\be
\frac{2 \pi}{\Gamma}\frac{d \Gamma}{d \phi} -
        \frac{2 \pi}{\bar \Gamma}\frac{d \bar \Gamma}{d \phi}=
- (\alpha_2 - \bar \alpha_2) cos2\phi - (\beta_2 - \bar \beta_2)sin2\phi
\ee
Other terms can be isolated by examining the $\phi$ dependence of the
differential rate difference between same hemisphere (SH) events
(e.g. $0<\theta_1,\theta_2<\frac{\pi}{2}$)
and opposite hemisphere (OH) events (e.g. $0<\theta_1<\frac{\pi}{2}$,
$\frac{\pi}{2}<\theta_2 < \pi$):
\be
\frac{2 \pi}{\Gamma}(\frac{d \Gamma^{OH}}{d \phi}
 - \frac {d \Gamma^{SH}}{d \phi})
 -\frac{2\pi}{\bar \Gamma}( \frac{d \bar \Gamma^{OH}}{d \phi} - \frac {d \bar
 \Gamma^{SH}}{d \phi}) =
 -\frac {1}{2}\{(\alpha_1 - \bar \alpha_1)cos \phi-(\beta_1 -\bar \beta_1)
  sin \phi\} \ee

In general the dominant terms in the angular correlations
 are $\Gamma_T/\Gamma$,
$\Gamma_L/\Gamma$, $\alpha_1$ and $\alpha_2$. The terms $\beta_1$
and $\beta_2$ are small and they are nonvanishing only if the helicity
amplitudes $H_{+1}, H_{-1}$ and $H_0$ or the invariant amplitudes $a, b$
and $c$, respectively, have different phases.
In all those channels where factorization is possible only in one way,
this is not the case because all matrix elements become simply
proportional to each other; this is, of course, due to the simplicity of
our model. For the same reason one has $\a_i = \bar{\a}_i$ (and $\G_T/\G =
\bar{\G}_T/\bar{\G}$) in all the channels with vanishing $\b_i$:
The overall (weak and/or strong) phase factor cancels in the ratios
that enter in the definition of the $\a_i$ [see (\ref{parameters})].

While the next-to-leading logarithmic precision of the effective
Hamiltonian allows one to consistently calculate all amplitudes at order
$\alpha_s$ and to include all one-loop matrix elements, some care is
necessary when evaluating CP-violating asymmetries of the decay rates
or of the observables of the angular distribution.
In particular, one should make sure that the rate asymmetries for
sufficiently inclusive channels remain consistent with CPT constraints
in certain mass limits \cite{GH}.

In order to specify a procedure which meets this requirement,
we recall that absorptive parts, which arise from intermediate
states having the {\it same} quark-gluon content as the final
state, play a crucial role
for CPT consistency of rate asymmetries in {\it inclusive} decays:
It has been shown (see e.g. \cite{Wolf,SEW}) that the inclusive rate
difference due to absorptive parts from these ``flavour-diagonal''
interactions must, and in fact does, cancel if all interferences
are taken into account which contribute at a given order in $\a_s$
(see appendix~B for an example of such a ``CPT cancellation'').
Of course, this cancellation need not be generally true when the
phase space of the
final state is restricted or when exclusive decays are considered.
Nevertheless, in our model for the {\it exclusive} amplitudes these
CPT cancellations hold for many of the diagrams due to the
factorization of the hadronic matrix elements. We shall assume that
analogous CPT cancellations are --- at least approximately --- valid for
all interferences of exclusive amplitudes to be evaluated here, and we
therefore neglect absorptive parts from flavor diagonal rescattering
throughout.

In this approximation all imaginary parts of the terms
$\Delta F_1(k^2/m_q^2)$ are to be dropped in (\ref{ct}) and (\ref{cp})
when $q$ refers to the flavor of a
$q\bar{q}$-pair which is present in the final state.
Moreover, no absorptive parts of vertex correction diagrams have to be
evaluated, because they are always flavour diagonal.
For consistency, the same procedure as for the rates should, of course,
be systematically applied when evaluating the decay parameters
(\ref{parameters}) and their CP-differences.
To get an idea of the quality of our approximation, we calculated the
flavour-diagonal absorptive parts for the case of
all penguin-like matrix elements, (\ref{ct}) and (\ref{cp}), and we
found that their effect is indeed small.

In our calculation we do not explicitly drop higher order terms
which arise, for instance, through interferences among (real and imaginary
parts of) the order $\a_s$ matrix elements.
However, such terms can not introduce the above mentioned inconsistencies
with CPT because the flavour-diagonal absorptive parts are discarded.
A completely systematic treatment of the higher order terms, some details
of which we describe in appendix~C, would require to expand all
{\it products} of interfering amplitudes --- and not the amplitudes
themselves --- in term of $\a_s$ and the other couplings of the
(effective) theory.

\section{Results and Discussion}

For a numerical analysis of the decay parameters and their CP-violating
effects within our model, we need to specify the CKM  matrix elements and
the current form factors. It is well known
\cite{schu} that fits for the parameters \footnote{
These coincide with the parameters $\rho$ and $\eta$ of the Wolfenstein
representation for small angles \cite{wolf1}. }
\begin{eqnarray}
\rho & = & cos \delta_{13}\; s_{13}/(s_{12}s_{23)}\nonumber \\
\eta & = & sin \delta_{13}\; s_{13}/(s_{12}s_{23)}
\end{eqnarray}
of the CKM matrix depend critically on the value of the $B$-meson decay
constant $f_B$. The solution for lower $f_B$ values leads to a negative
$\rho$ while higher $f_B$ values render $\rho$ positive.

We have calculated our results for the positive $\rho$ solution, with the
values
\bd \rho = 0.32\ ,\  \ \ \eta = 0.31\ed
(i.e. $s_{13} =0.0045,~ \delta_{13} = 44\deg$)
from the analysis by Schmidtler and Schubert\cite{schu} for $f_B = 250$ MeV
(giving $m_t = 135\pm 27$ GeV).
A more recent analysis by Ali and London \cite{Ali} based on the latest
information on $V_{ub}$ yields similar results. For comparison, we will
also show the asymmetries calculated with the negative $\rho$ solution:
$\rho = -0.41$ and $\eta =0.18$ \cite{schu} corresponding to $f_B = 125$
MeV (giving $m_t = 172\pm 15$ GeV).

The main purpose of this work is to calculate the effect of
(perturbative) strong phases on the angular correlation coefficients
obtained earlier with the strong phase put to zero \cite{KP}.
Moreover, we have included pure penguin channels and calculated partial
rate asymmetries. (The pure penguin modes are $B^- \to \rho^- K^{*0}$,
$K^{*-} \Phi$, $\rho^- \Phi$ and $B^-\to K^{*-} K^{*0}$.)
Of course, the estimates of the CP-violating observables given here
may suffer from large uncertainties due to strong phases from soft
final state interactions and, therefore, can at most be indicative of
the expected orders of magnitude.

First we consider the results for
the $\rho$ positive case (tables 1 and 2).
To see the effect of strong phases we also show in the following tables
the results without strong phases from imaginary parts of the
one-loop matrix elements (the values are given in parentheses and only
where different from those with strong phases included). These numbers
may differ from the results in \cite{Val,KP} since we include here
throughout the order $\a_s$ real parts of the penguin-like one-loop
matrix elements.
The branching ratios are calculated using $\tau_B = 1.49\:$p\,sec, and
a dash (---) in the tables indicates values which must
be exactly zero in our model.

The parameters $\b_1$ and $\b_2$ of the azimuthal decay distribution are
only non-zero for the decays $B^- \to K^{*-} \omega$, $K^{*-} \rho^0$ and
$\rho^- \omega$.
The other decays have $\b_i = 0$ in our model because all
three helicity amplitudes have the same overall (combined weak and strong)
phase. This is always the case when there is only one way to factorize
the hadronic matrix element, e.g. in $B^-\to K^{*-} \JPsi$, $D^{*-}_s
D^{*0}$, etc., and in all pure penguin modes. A special case is $B^- \to
\rho^0 \rho^-$: if isospin breaking (due to the mass difference
between $\rho^-$ and $\rho^0$) is neglected, no penguin contributions are
present\footnote{This has been missed in \cite{KP}.}
and, hence, no weak phase differences can occur.
Concerning the CP-violating effects in the angular distributions
the most promising decays in the case of vanishing strong phases
(values in parentheses) are
$B^-\to K^{*-} \omega$ and $B^-\to K^{*-}\rho^0$: They have branching ratios
of the order (1--3)$\times 10^{-6}$ and $\vert \b_1\vert \approx$(1--4)$\times
10^{-2}$ whereas the $\b_2$ are a factor of ten smaller. (See table 1.)

Table 1 also illustrates the influence of the strong phases generated by the
absorptive parts of the matrix elements. They have two effects: First,
they produce the rate asymmetry $a_{\rm CP} \not= 0$, which is given
in the third column, and second, they generate different angular
distributions for certain charge-conjugate decays, i.e.
$\a_i\not=\bar\a_i$ and $\b_i\not=\bar\b_i$.

The rate asymmetry $a_{\rm CP}$ is appreciable in
some of the cases, e.g. for $B^- \to K^{*-}  \omega$ ($a_{\rm CP}\approx
28\%$) and $B^- \to K^{*-} \rho^0$ ($a_{\rm CP}\approx 15\%$).
Both decays have approximately the same branching ratio of the order
(2--4)$\times 10^{-6}$. For the pure penguin modes these asymmetries
are either smaller ($\approx 1\%$) or, as in the case of
$B^- \to K^{*-}K^{*0}$, the branching ratios are tiny ($\sim 10^{-7}$).
Interesting is the decay $B^- \to D^{*-} D^{*0}$ with a branching ratio
of $0.1\%$ and a rate asymmetry\footnote{
This channel is rather insensitive to the particular choice of
$k^2$ because the leading absorptive part comes from a $u\ub$-cut with a low
threshold.} of about $1\%$.
For $B^- \to \rho^- K^{*0}$ the branching ratio and $a_{\rm CP}$ are
similar to the results for $\pi^-K^{*0}$ obtained recently by Fleischer
\cite{Flei}

For $\b_1$ and $\b_2$ we observe a significant effect from the
strong phase shifts for the most interesting
case $B^- \to  K^{*-}\omega$. From our results for $\bar{\b}_1$ and
$\bar{\b}_2$ for the charge conjugate decay $B^+ \to K^{*+}\omega$
(see the corresponding line in table~1) we find that $\b_i+\bar{\b}_i$
is not drastically changed  as compared to the case with no strong phases.
We obtained from table~1 $(\b_1+\bar{\b}_1)/2 = - 29\times 10^{-3}$ and
$(\b_2+\bar{\b}_2)/2 = 2.8\times 10^{-3}$ which is to be compared to
$\beta_1=\bar\b_1= -37\times 10^{-3}$ and $\beta_2=\bar\b_2=3.6\times 10^{-3}$.
On the other hand, we find $(\b_1-\bar{\b}_1)/2=-14\times 10^{-3}$ and
$(\b_2-\bar{\b}_2)/2=1.4\times 10^{-3}$ which shows the effect of the strong
phases [see eq. (16) and eq.(17)].
In the case of $K^{*-}\rho^0$ the behavior is less significant,
e.g. $(\b_1+\bar{\b}_1)/2 = 6.7\times 10^{-3}$ and
$(\b_1-\bar{\b}_1)/2 = -0.4\times 10^{-3}$ (compared to $\beta_1=\bar\b_1=
7.2\times 10^{-3}$ without strong phases).
In the $\rho \omega$ channel the pattern is even more pronounced: while
the $\b_i+\bar{\b}_i$ are again not changed substantially by the
strong phases, the differences,
$(\b_1-\bar{\b}_1)/2 = 0.16 \times 10^{-3}$ and $(\b_2-\bar{\b}_2)/2 =
-0.002\times 10^{-3}$, are now larger than the sums.

In the analogous calculations for $\a_i - \bar{\a}_i$
(and $\G_T/\G -\bar{\G}_T/\bar{\G}$) we find significant C-differences
only in the case of $K^{*-}\omega$, where the relative differences
of $\G_T/\G$, $\a_1$ and $\a_2$ are of the order of several percent
(detailed numbers can be extracted from table~1). In the two other cases,
$K^{*-}\rho^0$ and $\rho^- \omega$, these differences are smaller.

We have also performed the corresponding calculation with the $1/N$ terms
included throughout (see table~2). As a result, some of the predicted
branching ratios (BR) change drastically; for instance,
BR($B^- \to K^{*-}\JPsi)$
is decreased from $3.8\times 10^{-3}$ to $1.6\times 10^{-4}$.
Since the average experimental value of this branching ratio is
$(0.17\pm0.05)\%$~\cite{Stone} the version without $1/N$ terms is to
be preferred in this decay.
The pure penguin modes of the $b\to s$ transitions
are rather insensitive to the inclusion of the $1/N$ terms with the
exception of $K^*\Phi$ whose rate is increased by a factor of three.
The branching ratios for all other $b\to s$ transitions in the upper part
of the tables remain more or less unchanged, while $\b_1$ for $B^-\to
K^{*-}\omega$ and $B^-\to K^{*-}\rho^0$ become smaller and look less
interesting.
The pattern for $b\to d$ transitions (see the lower part of table~2)
is different when compared with table~1: the branching ratio
 for the pure penguin mode
$B^-\to \rho^-\Phi$ is drastically reduced and also the rate for
$B^-\to\rho^-\JPsi$ becomes more than an order of magnitude smaller.

It is obvious that the color unsuppressed decays $B^- \to D^{*-}_s
D^{*0}$, $D^{*0} D^{*-}$ (with amplitudes proportional to $\cb_2+\cb_1/N$)
are not significantly influenced by the treatment of the $1/N$ terms.
We should also note that the
coefficients $\Gamma_T/\Gamma$, $\Gamma_L/\Gamma$, $\alpha_1$ and
$\alpha_2$, in the angular distribution are not very
sensitive to the treatment of the $1/N$ terms and to
assumptions about the strong and weak phases. They depend mainly on
the helicity structure of the matrix elements and they are therefore
more important for testing the underlying model assumptions, in
particular, the current matrix elements in the BSW model in conjunction
with the factorization hypothesis.

To give an impression how sensitive our results are with respect to the
solution ambiguity in the CKM parameter determination we have repeated
the calculations with $\rho$ negative. The results
(without 1/N terms) are shown in table~3 and should be compared
with the numbers in table~1 (with strong phases).
Generally the branching ratios and
asymmetries are similar in magnitude. In the interesting $K^{*-}\omega$
and $K^{*-}\rho^0$ final states the branching ratios decrease or
increase, respectively, while the rate asymmetries and azimuthal
asymmetries vary in the opposite direction. The pure penguin modes
change less but in a similar way. Since the charge asymmetries of
the various observables can be more pronounced for a $\rho$ negative
CKM matrix, the interesting charge conjugate channels are also
included in table~3.

We mention that some of the pure penguin modes have been calculated by
other authors. Davies et al. obtained a comparable result for the
partial rate of $B \rightarrow K^*\Phi$ \cite{Davies} using only leading
logarithmic order QCD coefficients. Dong-sheng Du et al. \cite{Du}
calculated the $\rho \Phi $ rate using also coefficients of Buras et al..
They obtained similar results, in particular also the strong suppression
of this rate when 1/N terms are included.

In this work we have confined our attention to direct CP violation in
decays of charged $B$ mesons. The strong phases also contribute to
some {\it neutral} $B$ decays, where their effect on the CP-violating
time dependence is a further complication in the analysis of
CP violation arising from the interference of mixing and decay amplitudes.
This subject is currently under investigation.

Finally, we would like to note that the partial rate asymmetries we have
found are larger than those reported in $B$ decays to two pseudoscalar
mesons, making $B \rightarrow VV $ an attractive channel for probes of
direct CP violation.

\section*{Acknowledgement}
W.\,F.\,P. thanks the Desy Theory Group for its kind hospitality and the North
Atlantic Treaty Organization for a Travel Grant. G.~K. thanks the Department of
Physics of OSU for hospitality and financial support of his visit. We like to
thank D.~Wyler for helpful comments on the manuscript.

\newpage
\subsection*{Appendix A: Matrix elements}
The momentum dependence of the penguin-like matrix elements entering in
(\ref{ct}) and (\ref{cp}) is given by
\bea
\D F_1 (z) & = & -4 \int^1_0 u(1-u) \ell n \left[1-z u(1-u)\right] du
\nonumber \\[2mm]
& = & \frac{2}{3} \left\{ \frac{5}{3} + \frac{4}{z} +
      \bigl(1+\frac{2}{z}\bigr) R(z) \right\} \ ,\eea
where, setting $r\equiv \sqrt{|1-4/z|}$,
\bd
R(z) = \left\{ \ba{ll}
r\cdot \ln\frac{r-1}{r+1}               & (z<0) \\[1mm]
-2 + \frac{z}{6} + \frac{z^2}{60} + \frac{z^3}{420} + \cdots   & (z\to 0)
\\[1mm]
 - r\pi  + 2r {\rm arctan}\; r         & (0<z<4) \\[1mm]
+  \i r\, \pi + r \ln\frac{1-r}{1+r} & (z>4) \ea \right. \ed
\\

For completness, we list here also the factorized matrix elements of
$\Heff$ for the pure penguin modes not yet presented in former publications
\cite{KP}:
\def\N{{{\displaystyle \frac{1}{N}}}}
\bea
\bra \rho^- K^{*0} \vert \Heff^{(q)} \vert B^- \ket & = &
v_q \sqrt{2} G_F
\Bigl( \N\ce_3 + \ce_4\Bigr)
\bra \rho^- \vert \db \g_\m b_L \vert B^- \ket
\bra K^{*0} \vert \sb \g^\m d   \vert 0 \ket \ ,\\
\bra K^{*-} \Phi \vert \Heff^{(q)} \vert B^- \ket & = &
v_q \sqrt{2} G_F
\Bigl( (1+\N)\ce_3 + (1+\N) \ce_4 + \ce_5 + \N \ce_6 \Bigr)
\nonumber \\ && \cdot
\bra K^{*-} \vert \sb \g_\m b_L \vert B^- \ket
\bra \Phi \vert \sb \g^\m s \vert 0 \ket \ ,\\
\bra \rho^- \Phi \vert \Heff^{(q)} \vert B^- \ket & = &
v_q \sqrt{2} G_F
\Bigl( \ce_3 + \N \ce_4 + \ce_5 + \N \ce_6 \Bigr)
\nonumber \\ && \cdot
\bra \rho^- \vert \db \g_\m b_L \vert B^- \ket
\bra \Phi \vert \sb \g^\m s \vert 0 \ket \ ,\\
\bra K^{*-} K^{*0} \vert \Heff^{(q)} \vert B^- \ket & = &
v_q \sqrt{2} G_F
\Bigl( \N\ce_3 + \ce_4 \Bigr)
\bra K^{*-} \vert \sb \g_\m b_L \vert B^- \ket
\bra K^{*0} \vert \db \g^\m s \vert 0 \ket \ .
\eea

\subsection*{Appendix B: CPT cancellations in inclusive rate asymmetries}
In this appendix we shall illustrate a typical example of a ``CPT
cancellation'' for the CP-violating rate difference in inclusive decays.
In contrast to the discussion within the full theory of refs.~\cite{GH,SEW}
we adopt here the framework of the effective Hamiltonian where one
can perform a completely analogous analysis.
To render the correspondence to diagrams in the full theory more obvious,
we assume the simplified situation where all $\cb_i$ are zero,
except $\cb_2$.

An illustrating channel is the charmless $\Delta b=\Delta s=1$ transition
$b\to s u \bar u$: At order $\alpha_s$ the rate difference
$a_{CP} \sim \G-\bar\G$
can arise only from the interference between the tree-level matrix element
of $O_2^{(u)}$ and the penguin diagram of $O_2^{(c)}$ (see fig.~2).
The resulting asymmetry is small because the absorptive part of the
penguin is kinematically suppressed by the $c\cb$-threshold. At order
$\alpha_s^2$ one finds two types of contributions: First, the
interference between penguin-like matrix elements of $O_2^{(u)}$ and
$O_2^{(c)}$ (see fig.~3a), and second, interferences between the tree
level matrix element of $O_2^{(u)}$ and order $g_s^4$ diagrams with an
insertion of $O^{(c)}_2$. In particular, there is the
interference with a penguin-like diagram of $O_2^{(c)}$ having an
additional vacuum polarization on the gluon line (fig.~3b). An
absorptive part of this diagram can be generated through a $u\ub$-pair
inside the loop. In this case, the two order $\alpha_s^2$
interferences depicted in fig.~3 differ only by interchanging the role
of the final and the ``cut'' state, which generates the absorptive part.
Applying the Cutkosky rules one readily finds (see also \cite{Wolf,SEW})
that both interferences have the same size, and their combined effect
cancels due to the relative minus sign in
\bd \G - \bar\G \sim \Im[A_u A_c^*] =
\Im[A_u]\cdot\Re[A_c] - \Re[A_u]\cdot\Im[A_c]\ . \ed
Note that the cancellation between the diagrams of, for instance, fig.~3
works also for the (factorized!) {\it exclusive} amplitudes because
the quark loop contributes in both cases the same multiplicative factor.

\subsection*{Appendix C: Systematic expansion of the observables}
In the framework of the effective Hamiltonian described in section~2,
the natural choice for the couplings, in terms of which the observables
are to be expanded, is $\a_s$ at $\mu = m_b$ together with the
manifestly renormalization scheme independent coefficients $\cb_i$.
Of course, CPT cancellations analogous to the ones discussed above
are present for each interference proportional to $\cb_i \cb_j \a_s^n$
with $n\ge 2$.
In table~4 we have specified the order up to which our calculation of the
various observables is complete.
Here, we use $c_t$ and $c_p$ as generic notations for any of
the coefficients $\{\cb_1,\cb_2\}$ of $O_{1,2}$ or of the coefficients
$\{\cb_3,\ldots,\cb_6\}$ of the penguin operators, respectively
[not necessarily in the combinations defined in (\ref{ct}) and (\ref{cp})].

For decay modes with tree and penguin contributions we do not cover any
terms of order $\a_s^2$ in a complete way.
However, since we are counting the powers of $\alpha_s(m_b)$ within
the {\it effective} theory, the interference between local penguin
contributions ($\sim c_p$) and absorptive penguin-like matrix elements
($\sim \a_s c_t$) is included here already at order $\a_s$ (while it is
order $\alpha_s^2$ in the full theory). For instance,
the local parts of the subdiagrams in fig.~3 (viewed as diagrams
in the full theory), which are enclosed by a box, are taken into account
in the effective theory through the penguin operators $O_3,\ldots,O_6$.
Although the coefficients $c_p$, being implicitly of order $\alpha_s(M_W)$,
are small, the order $\a_s c_t c_p$ interferences can be numerically
important for the rate asymmetries in some charmless decays
where the  $\a_s c_t^2$ contributions (see e.g. Fig.~2) are kinematically
suppressed due to the $c\cb$-threshold \cite{GH}.

In the case of pure penguin modes, i.e. final states which have at tree
level only contributions from the penguin operators $O_3,\ldots,O_6$,
interferences of order $\alpha_s^2 c_t^2$ can arise only as a product
of two order $g_s^2$ amplitudes.
Therefore, the penguin-like one-loop matrix elements are sufficient for
the complete treatment of these interferences. On the other hand,
we may neglect systematically all terms of order $\a_s c_p^2$ and
$\a_s^2 c_t c_p$.
Thus, we retain effectively the terms which originate from
order $\alpha_s^2$ contributions in the full theory (and before the
renormalization group evolution which sums up powers of $\a_s\times \ln
\mu^2/M_W^2$).
Of course, this procedure is also numerically sensible, because
$c_t \gg c_p$ by at least a factor of five ($\approx 1/\alpha_s$)
due to their origin from penguin diagrams.

At any order higher than the ones listed in table~4, real parts of
vertex-correction diagrams are necessary for a complete treatment.
Their imaginary parts, which enter already in the terms enclosed
by parentheses in the table, correspond to flavour-diagonal rescattering
and are neglected throughout our calculations.
Of course, a complete treatment of all order $\a_s^2$ interferences
also requires one to calculate a large number of two-loop diagrams \cite{SEW}
for the matrix elements of $\Heff$. In fact, the next-to-leading logarithmic
precision for the coefficient functions, just allows one to evaluate their
absorptive parts (but not their real parts) without encountering new
renormalization scheme dependencies.

\newpage
\section*{Figure captions}
\begin{description}

\item [Fig.\,1:] Two types of one-loop matrix elements: (a) Vertex
corrections, and (b) penguin diagrams. The square box denotes an
insertion of one of the four-quark operators $O_i$ of \eq(\ref{operators}).

\item [Fig.\,2:] Two interfering diagrams generating a rate asymmetry at
order $\a_s$. The absorptive phase arises from the cut state indicated
by the dashed line.

\item [Fig.\,3:] Examples of order $\a_s^2$ interferences: (a) Two
$O(g_s^2)$ penguin diagrams, and (b) a tree and a $O(g_s^4)$ diagram.
The rate differences due to the absorptive parts from the $\ub u$ cut
(dashed line) cancel. The local part of the subdiagrams enclosed by
the dotted box is taken into account by penguin operators in the
effective theory.
\end{description}

\section*{Table captions}
\begin{description}
\item [Tab.\,1:] Branching ratios, rate asymmetries and angular correlation
coefficients, using matrix elements {\it without} $1/N$ Terms
for a $\rho$ {\it positive} CKM Matrix ($\rho=0.32, \eta=0.31$).
Values in parentheses correspond to the case without strong phases.

\item [Tab.\,2:] Branching ratios, rate asymmetries and angular correlation
coefficients, using matrix elements {\it with} $1/N$ Terms
for a $\rho$ {\it positive} CKM Matrix ($\rho=0.32, \eta=0.31$).
Values in parentheses correspond to the case without strong phases.

\item [Tab.\,3:] Branching ratios, rate asymmetries and angular correlation
coefficients, using matrix elements {\it without} $1/N$ Terms
for a $\rho$ {\it negative} CKM Matrix ($\rho=-0.41, \eta=0.18$).

\item [Tab.\,4:] Orders of $\a_s$, $c_t$ and $c_p$ at which the
treatment of the various observables is complete (neglecting
absorptive parts from flavour-diagonal rescattering in
vertex-correction diagrams).
\end{description}

\newpage
\renewcommand{\baselinestretch}{1.0}

\def\eleven{{B^- \to K^{*-} + \JPsi}}
\def\twelve{{ B^- \to D_s^{*-} + D^{*0}}}
\def\fifteen{{B^- \to \rho^- + \omega }}
\def\fifteenbar{{B^+ \to \rho^+ + \omega }}
\def\sixteen{{B^- \to \rho^- + \rho^0}}
\def\eighteen{{B^- \to \rho^- + \JPsi}}
\def\twentyseven{{B^-\to K^{*-} + \omega}}
\def\twentysevenbar{{B^+\to K^{*+} + \omega}}
\def\twentyeight{{B^-\to K^{*-} + \rho^0}}
\def\twentyeightbar{{B^+\to K^{*+} + \rho^0}}
\def\twentynine{{B^-\to D^{*0} + D^{*-}}}
\def\thirtyeight{{B^-\to K^{*-} + \Phi}}
\def\thirtynine{{B^-\to K^{*-} + K^{*0}}}
\def\forty{{B^- \to \rho^- + \Phi}}
\def\fortyone{{B^- \to \rho^- + K^{*0}}}

\def\tmtwo{$\times10^{-2}$}
\def\tmthree{$\times10^{-3}$}
\def\tmfour{$\times10^{-4}$}
\def\tmfive{$\times10^{-5}$}
\def\tmsix{$\times10^{-6}$}
\def\tmseven{$\times10^{-7}$}
\def\tmeight{$\times10^{-8}$}

\renewcommand{\arraystretch}{1.3}
\def\nc{\\}
\def\sc{\\[-1mm]}
\small

\def\empty{ (---) &&&&& }
\def\same{ }

\begin{center}
Table 1 \\[1cm]
\bigskip
\begin{tabular}{||l|c|c|c|c|c|c|c||}
\hline\hline
\multicolumn{8}{||c||}{Matrix Elements {\it without} $1/N$ Terms
and {\it with (without)} Strong Phases}\\
\multicolumn{8}{||c||}{$\rho~positive$ CKM Matrix: $\rho=0.32, \eta=0.31$}\\
\hline
Channel & BR & $a_{CP}$ & $\G_T \over \G$ & $\a_1$ & $\a_2$ &
$\b_1$ & $\b_2$ \\
        &    &    [\%]  &                 &        &        &
$[10^{-3}]$ & $[10^{-3}]$ \\[1mm]
\hline
\multicolumn{8}{||c||}{$b \rightarrow s$ transitions:~~
$\Delta c=0,~\Delta b =  \Delta s = 1 $}\\
\hline
$\fortyone$   & 1.3\tmfive & 0.54  & 0.107 & $-$0.334 & 0.009 & --- & ---   \sc
              &(1.2\tmfive ) & \empty \nc
$\thirtyeight$ & 5.5\tmsix & 1.22  & 0.137 & $-$0.385 & 0.017 & --- & --- \sc
             & (4.3\tmsix) & \empty \nc
$\twentyseven$   & 2.4\tmsix & 28 & 0.089 & $-$0.316 & 0.011 & $-$44 & 4.2\sc
$\twentysevenbar$& 1.3\tmsix &    & 0.073 & $-$0.300 & 0.012 & $-$15 & 1.5\sc
   & (1.7\tmsix  ) & (---) & (0.080) & ($-0.307$)&(0.012)&($-$37)&(3.6)\nc
$\twentyeight$   & 3.7\tmsix & 15 & 0.105 & $-$0.332 & 0.009 & 7.1&$-$0.63\sc
$\twentyeightbar$& 2.7\tmsix &    & 0.103 & $-$0.330 & 0.009 & 6.3&$-$0.55\sc
   & (3.0\tmsix  ) & (---)  & (0.104) & ($-$0.331) & (0.009)&(7.2)&($-$0.64)\nc
$\eleven$      & 3.8\tmthree &  --- & 0.427 & $-$0.621 & 0.123 & --- & ---\sc
              & (3.8\tmthree)& \empty \nc
$\twelve$      & 2.4\tmtwo  & $-$0.05& 0.476 & $-$0.664 & 0.183 & --- & ---\sc
              & (2.4\tmtwo  ) & \empty \nc
\hline
\multicolumn{8}{||c||}{$b \rightarrow d$ transitions:~~
$\Delta s = \Delta c = 0,~\Delta b = 1 $}\nc
\hline
$\forty$      & 1.1\tmseven &  --- & 0.130 & $-$0.364 & 0.011 & --- & --- \sc
            & (1.1\tmseven) & \empty \nc
$\thirtynine$ & 4.2\tmseven & $-18$ & 0.112 & $-0.352$ & 0.014 & --- & ---\sc
            & (4.8\tmseven) & \empty \nc
$\fifteen$   & 1.3\tmfive & $-6.6$ & 0.083 & $-0.298$ & 0.007
&$+0.22$&$-0.003$\sc
$\fifteenbar$& 1.5\tmfive &        & 0.083 & $-0.298$ & 0.007
&$-0.10$&$+0.001$\sc
 & (1.4\tmfive ) & (---) & \same & \same & \same &($0.05$) & ($-0.001$)\nc
$\sixteen$    & 1.3\tmfive  &  --- & 0.083 & $-0.298$ & 0.007 & --- & --- \sc
            & (1.3\tmfive ) &  \empty \nc
$\eighteen$   & 1.8\tmfour  &  --- & 0.388 & $-0.597$ & 0.097 & --- & ---\sc
             & (1.8\tmfour ) & \empty \nc
$\twentynine$ & 1.2\tmthree & 0.89  & 0.456 & $-0.660$ & 0.172 & --- & ---\sc
            & (1.2\tmthree) & \empty \nc
\hline\hline
\end{tabular}
\end{center}
\eject

\begin{center}
Table 2 \\[1cm]
\bigskip
\begin{tabular}{||l|c|c|c|c|c|c|c||}
\hline\hline
\multicolumn{8}{||c||}{Matrix Elements {\it with} $1/N$ Terms
and {\it with (without)} Strong Phases}\\
\multicolumn{8}{||c||}{$\rho~positive$ CKM Matrix: $\rho=0.32, \eta=0.31$}\\
\hline
Channel & BR & $a_{CP}$ & $\G_T \over \G$ & $\a_1$ & $\a_2$ &
$\b_1$ & $\b_2$ \\
        &    &    [\%]  &                 &        &        &
$[10^{-3}]$ & $[10^{-3}]$ \\[1mm]
\hline
\multicolumn{8}{||c||}{$b \rightarrow s$ transitions:~~
$\Delta c=0,~\Delta b =  \Delta s = 1 $}\\
\hline
$\fortyone$    & 1.0\tmfive& 0.56 & 0.107 & $-0.334$ & 0.009 & ---   & --- \sc
             & (9.3\tmsix) & \empty \nc
$\thirtyeight$ & 1.5\tmfive& 0.56 & 0.137 & $-0.385$ & 0.017  & ---  & --- \sc
               &(1.4\tmfive) & \empty \nc
$\twentyseven$ & 2.6\tmsix & 29 & 0.107 & $-0.334$ & 0.009 & $-0.95$ & 0.09\sc
        & (1.8\tmsix)&(---) & (0.107) &$(-0.333)$  &(0.009)&($-1.2$)&(0.12)\nc
$\twentyeight$ & 2.5\tmsix & 30 & 0.106 & $-0.333$ & 0.009 & $-1.6$ & 0.14\sc
        & (1.8\tmsix)&(---)&(0.106)&($-0.333$)&(0.009)&($-1.9$)&(0.17)\nc
$\eleven$      & 1.6\tmfour & --- & 0.427 & $-0.621$ & 0.123 & ---  & ---\sc
             & (1.6\tmfour) & \empty \nc
$\twelve$     & 2.0\tmtwo   &$-0.05$& 0.476 & $-0.664$ & 0.183 & ---  & ---\sc
             & (2.0\tmtwo ) & \empty \nc
\hline
\multicolumn{8}{||c||}{$b \rightarrow d$ transitions:~~
$\Delta s = \Delta c = 0,~\Delta b = 1 $}\\
\hline
$\forty$       & 1.6$\times 10^{-11}$& --- & 0.130 & $-0.364$ & 0.011
& --- & --- \sc
           & (1.6$\times 10^{-11}$)& \empty \nc
$\thirtynine$  & 3.1\tmseven &$-18$ & 0.112 & $-0.352$ & 0.014 & --- & ---\sc
             & (3.6\tmseven) & \empty \nc
$\fifteen$     & 2.5\tmfive  &$-4.1$ & 0.084 &$-0.299$& 0.007&0.23&$-0.003$\sc
             & (2.6\tmfive ) &(---) &\same &\same &\same &(0.17)&($-0.002$)\nc
$\sixteen$     & 2.3\tmfive  & --- & 0.083 & $-0.298$ & 0.007 & --- & --- \sc
             & (2.3\tmfive ) & \empty \nc
$\eighteen$    & 7.0\tmsix   & --- & 0.388 & $-0.597$ & 0.097 & --- & ---\sc
             & (7.0\tmsix  ) & \empty \nc
$\twentynine$  & 1.0\tmthree & 0.88 & 0.456 & $-0.660$ & 0.172 & --- & ---\sc
             & (1.0\tmthree) & \empty \nc
\hline\hline
\end{tabular}
\end{center}

\newpage

\begin{center}
Table 3 \\[1cm]
\bigskip
\begin{tabular}{||l|c|c|c|c|c|c|c||}
\hline\hline
\multicolumn{8}{||c||}{Matrix Elements {\it without} $1/N$ Terms
and {\it with} Strong Phases}\\
\multicolumn{8}{||c||}{$\rho~negative$ CKM Matrix: $\rho=-0.41, \eta=0.18$}\\
\hline
Channel & BR & $a_{CP}$ & $\G_T \over \G$ & $\a_1$ & $\a_2$ &
$\b_1$ & $\b_2$ \\
        &    &    [\%]  &                 &        &        &
$[10^{-3}]$ & $[10^{-3}]$ \\[1mm]
\hline
\multicolumn{8}{||c||}{$b \rightarrow s$ transitions:~~
$\Delta c=0,~\Delta b =  \Delta s = 1 $}\\
\hline
$\fortyone$    & 1.3\tmfive  & 0.33  & 0.107 & $-0.334$ & 0.009 & --- & ---
\sc
$\thirtyeight$ & 5.2\tmsix   & 0.75  & 0.137 & $-0.385$ & 0.017 & --- & --- \sc
$\twentyseven$  & 6.5\tmseven & 83 & 0.135 & $-0.342$ &$+0.004$ &$-131$
&$-13$\sc
$\twentysevenbar$&6.0\tmeight &    & 0.386 & $-0.402$ &$-0.047$ &$+231$
&$-22$\sc
$\twentyeight$   & 9.0\tmsix & 3.1  & 0.109 & $-0.335$ & 0.009
&$+0.74$&$-0.06$\sc
$\twentyeightbar$& 8.5\tmsix &      & 0.109 & $-0.335$ & 0.009
&$+2.18$&$-0.19$\sc
$\eleven$      & 3.9\tmthree &  --- & 0.427 & $-0.621$ & 0.123 & --- & ---\sc
$\twelve$      & 2.4\tmtwo  & $-0.03$ & 0.476 & $-0.664$ & 0.183 & --- & ---\nc
\hline
\multicolumn{8}{||c||}{$b \rightarrow d$ transitions:~~
$\Delta s = \Delta c = 0
,~\Delta b = 1 $}\\
\hline
$\forty$      & 3.9\tmseven &  --- & 0.130 & $-0.364$ & 0.011 & --- & ---- \sc
$\thirtynine$ & 1.4\tmsix   & $-3.5$ & 0.112 & $-0.352$ & 0.014 & --- & ---\sc
$\fifteen$    & 9.1\tmsix   & $-5.6$ & 0.083 & $-0.298$ & 0.007 &
$-0.12$&$+0.001$\sc
$\fifteenbar$ & 10.\tmsix   &        & 0.083 & $-0.298$ & 0.007 &
$+0.18$&$-0.002$\sc
$\sixteen$    & 1.3\tmfive  &  --- & 0.083 & $-0.298$ & 0.007 & --- & --- \sc
$\eighteen$   & 1.6\tmfour  &  --- & 0.388 & $-0.597$ & 0.097 & --- & ---\sc
$\twentynine$ & 1.2\tmthree & 0.55  & 0.456 & $-0.660$ & 0.172 & --- & ---\nc
\hline\hline
\end{tabular}
\end{center}

\vspace*{1cm}
\begin{center} Table 4\\
\vspace{0.6cm}
\begin{tabular}{|c|c|c|}
\hline
Observables  & \multicolumn{2}{|c|}{Decay mode}\\
             & tree \& penguin  & pure penguin \\
\hline
$\Gamma$, $\a_i$
   & $c_t^2$\,,\ $c_t c_p$\,,\ $c_p^2$
   & $c_p^2$\,,\ $\a_s c_t c_p$\,,\ $\a_s^2 c_t^2$ \\
$\G-\bar\G$\,,\ $\a_i-\bar\a_i$
   & $\a_s c_t^2$\,,\ $(\a_s c_t c_p$\,,\ $\a_s c_p^2)$
   & $\a_s c_t c_p$\,,\ $(\a_s c_p^2)$\,,\ $\a_s^2 c_t^2$ \\
$\b_i$
   & $c_t c_p$\,,\ $c_p^2$\,,\ $\a_s c_t^2$
   & $c_p^2$\,,\ $\a_s c_t c_p$\,,\ $\a_s^2 c_t^2$ \\
$\b_i-\bar\b_i$
   & $(\a_s c_t^2$\,,\ $\a_s c_t c_p$\,,\ $\a_s c_p^2)$
   & $\a_s c_t c_p$\,,\ $(\a_s c_p^2)$\,,\ $\a_s^2 c_t^2$\\
\hline
\end{tabular}
\end{center}

\end{document}